\documentclass[a4,aps,twocolumn,showpacs,preprintnumbers,amsmath,amssymb]{revtex4} 
\usepackage{dcolumn}
\usepackage{graphicx}
\usepackage{amsfonts} 
\usepackage{epstopdf}
\usepackage{color}
\usepackage{here}
\begin{document} 
\title{Stability of persistent currents in spinor Bose-Einstein condensates}
\date{\today}
\author{A.I. Yakimenko$^{1}$, K.O. Isaieva,$^1$, S.I. Vilchinskii$^1$, M. Weyrauch$^{2}$}
\affiliation{$^1$Department of Physics, Taras Shevchenko National University of Kyiv, 64/13, Volodymyrska Str. City of Kyiv, 01601, Ukraine \\
$^2$Physikalisch-Technische Bundesanstalt, Bundesallee 100, D-38116 Braunschweig, Germany}

\begin{abstract}
Motivated by a recent experiment [S. Beattie, S. Moulder, R. J. Fletcher, and Z. Hadzibabic, PRL 110, 025301 (2013)] we study the superflow of atomic spinor Bose-Einstein condensates optically trapped in a ring-shaped geometry. Within a dissipative mean-field approach we simulate a two-component condensate in conditions
adapted to the experiment. In qualitative agreement with the experimental findings, we observe persistent currents, if the spin-population imbalance is above some well-defined `critical' value. The triply charged vortices decay in quantized steps. The vortex lines escape from the center of the ring through dynamically created regions in the condensate annulus with reduced density of one component filled by atoms of the other component. The vortices then leave the ring-shaped high density region of the condensate and finally decay into elementary excitations.

\end{abstract}

\pacs{05.45.Yv, 03.75.Lm, 05.30.Jp} \maketitle

Persistent currents or `flows without friction' as a hallmark of superfluidity have been studied in liquid helium
for several decades. Recently, persistent flow of atoms has been observed
in Bose-Einstein condensates (BEC) trapped in a
ring-like potential~\cite{Ryu07,Ramanathan11,CavendishSingleComponent,Wright13,Beattie13}.
This enables fundamental studies of superfluidity and may lead to
applications in high precision metrology
and atomtronics. Of course, the question of the stability of the atomic persistent currents is of fundamental importance and, therefore, the subject of numerous investigations.

Theoretical studies of persistent currents in atomic BEC are mostly limited to the simplified cases of one-dimensionality and very weak interactions.
One-component BECs in a one-dimensional (1D) ring potential were studied in Refs.~\cite{KanamotoUedaPRA09} and \cite{KanamotoUedaPRA10}. Superfluidity on a 1D ring in the presence of impurities was investigated in Ref.~\cite{YulinPRA11}.
Families of 2D solitary waves with and without singly-charged persistent
flow are investigated in Ref.~\cite{Berloff09}.  For two-component BECs in 1D or 2D traps, the stability of the persistent currents and their decay mechanisms  were under investigation in Refs. \cite{SmyrnakisPRL09,SmyrnakisPRA10,SmyrnakisPRA10_annular,SmyrnakisPRA07,JacksonKavoulakisPRA06,KavoulakisJLTP07,BrtkaMalomedPRA10}.
Smyrnakis {\it et al.}~\cite{SmyrnakisPRL09} concluded that in a strictly one-dimensional ring persistent currents with  circulation lager than one  are stable only in single-component gases. This conclusion was challenged recently~\cite{ANO13}.
In Ref.  \cite{BargiKavoulakis10} it was found on the basis of mean field calculations supported by exact diagonalization results, that  persistent currents in 2D traps may be stable under specific conditions.

Experimentally, this problem has been addressed very recently~\cite{Beattie13} for a toroidally trapped gas of $^{87}$Rb atoms in two different spin states. However, previous theoretical investigations describe the stability of the persistent currents in spinor BEC only qualitatively, but do not elucidate the microscopic mechanism of the instabilities and its impact on the dynamics of the persistent currents.

In the present work, we investigate the stability of the superflow  within a two-dimensional dissipative mean-field theory. We find that  phase-slips occur as the result of the simultaneous action of two factors: an azimuthal symmetry-breaking instability and a dissipation effect caused by the interaction of the condensate atoms with the thermal cloud. If the number of atoms in the minor spin component becomes comparable to the number of atoms in the major component, the nonlinear repulsive inter-component interaction leads to a separation of the spin components. As a result of this instability, one finds regions in the condensate annulus with reduced density of one component filled by atoms of the other component.
These regions serve as `gates' for the vortex lines where they readily cross the annulus, and, as a consequence, the winding number of the superflow is reduced by one unit.

\textit{Model} --
The main properties of an ultracold dilute spin-1 atomic BEC at zero temperature can be accurately described by a set of mean-field Gross-Pitaevskii  equations (GPEs) for the spin components $(\Psi_+,\Psi_0,\Psi_-)$ of the three-component order parameter $\vec\Psi$. These equations correspond to the Hamiltonian $H=H_0+H_A$ with the spin-independent part (see, e.g. ~\cite{Matuszewski12})
\begin{equation}\label{hamilt1}
H_0=\sum_j\int \Psi_j^*\left(-\frac{\hbar^2}{2M}\Delta+\frac{c_0}{2}n+V(\textbf{r})\right)\Psi_j  d^3\textbf{r},
\end{equation}
where $\textbf{r}=(x,y,z)$ and $n=n_++n_0+n_-=\sum_j|\Psi_j|^2$ is the total density; the strength of interaction between atoms $c_0=\frac{4 }{3}\hbar^2\pi(2a_2+a_0)/M$ is given in terms of the s-wave scattering length $a_S$ for atom pairs with total spin $S$.
The spin-dependent part $H_A$ of the Hamiltonian is given by
\begin{equation}\label{hamilt2}
H_A=\int \left(E_+n_++E_-n_-+E_0n_0+\frac{c_2}{2}|\textbf{F}|^2\right) d^3\textbf{r}
\end{equation}
where $E_j$ is the Zeeman energy of the state $\Psi_j$, $\textbf{F}$ the spin density and $c_2=\frac{4 }{3}\hbar^2\pi(a_2-a_0)/M$.

A vortex line manifests itself by the emergence of a hole in the condensate, around which the atoms of  spin component $j$ rotate with
velocities $\textbf{v}_j=\frac{\hbar}{M}\nabla\Phi_j$, where $\Phi_j$ is the phase of the condensate wave function: $\Psi_j=|\Psi_j|e^{i\Phi_j}$, subject to the quantization condition
\begin{equation}\label{quantization}
M\oint_\Gamma {\textbf{v}}_jd\textbf{l}=2\pi \hbar q_j,
\end{equation}
where $\Gamma$ indicates a closed contour around the vortex core, and the integer $q_j$ is the topological charge of the spin component $j$. The persistent flow can be characterized by a single $q$-charged vortex line pinned at the center of the ring-shaped condensate.
The external ring-shaped trap produces a huge central hole at the axis of the condensate cloud, where the vortex energy has a local minimum, thus the vortex core in toroidal traps is bounded by the potential barrier, which makes even the giant multi-charged vortices robust.

In accordance with the experimental set-up of Ref.~\cite{Beattie13}, we approximate the external toroidal optical trapping potential $V(\textbf{r})$ by a superposition of a harmonic potential with trapping frequency $\omega_z$ (which models the elliptic highly anisotropic `sheet' beam) and a radial Laguerre-Gauss potential (which models the `tube beam'):
\begin{equation}
  V(\textbf{r})=\frac{M\omega_z^2 z^2}2-V_0\left(\frac{r}{R}\right)^{2q} e^{-q\left(r^2/R^2-1\right)},
\end{equation}
where $M$ is the atomic mass, $R$ is the radius of the trap (radial coordinate of the minimum of the trap), $r=\sqrt{x^2+y^2}$, and $V_0$ is the trap depth. 
The parameter $q=3$ corresponds to the topological charge of the optical vortex which produces the Laguerre-Gauss potential.

The harmonic potential creates a tight binding potential in $z$ direction, so that the BEC cloud is ``disk-shaped" ($l_z\ll R$ with $l_z=\sqrt{\hbar/(M\omega_{z})}$ the longitudinal oscillator length). Consequently, we assume that the longitudinal motion of condensate is frozen in, $\Psi_j(\mathbf{r},t)=\tilde\Psi_j(r,t)\Upsilon(z,t),$ where
$\Upsilon(z,t)=(l_{z}\sqrt{\pi})^{-1/2}\exp(-\frac{i}{2}\omega_zt-\frac12z^2/l_{z}^2)$. After integrating out the longitudinal coordinates in the  GPEs corresponding to the Hamiltonian $H$ given in Eq.~(\ref{hamilt1}) and (\ref{hamilt2}), and accounting for dissipative effects (see below), we obtain a  set of three coupled differential equations in 2D:
\begin{eqnarray}
  (i-\gamma)\frac{\partial \psi_\pm}{\partial
  t}&=&\hat{\mathcal{H}}_\pm\psi_\pm+\nu_a n_0\psi^*_\mp,\label{main1}\\
  (i-\gamma)\frac{\partial \psi_0}{\partial
  t}&=&\hat{\mathcal{H}}_0\psi_0+2\nu_a\psi_+\psi_-\psi^*_0,\label{main2}
\end{eqnarray}
with
\begin{eqnarray}
\hat{\mathcal{H}}_\pm&=&-\frac{1}{2}\Delta_\perp-\mu_\pm+V(r)+\nu_s n +\nu_a(n_0+n_\pm-n_\mp), \nonumber\\ \hat{\mathcal{H}}_0&=&-\frac{1}{2}\Delta_\perp-\mu_0+V(r) -\epsilon +c_0n +c_2(n_++n_-).\nonumber
\end{eqnarray}
Here,  we have introduced  the 2D Laplace operator $\Delta_\perp$ and dimensionless space and time coordinates $r\rightarrow r/R$, $z\rightarrow z/l_z$, $t\rightarrow \omega_r t$ with $\omega_r=\hbar/({M R^2})=4.8~\textrm{Hz}$ as well as the dimensionless chemical potential $\mu_j\to \mu_j/(\hbar\omega_r)$ of the  spin component $j$.   The order parameter takes the form $\psi_j e^{-i\mu_j t}=\tilde\Psi_j/C$ with $C^2=\sqrt{2\pi}\hbar\omega_r l_z/c_0$.
Furthermore we define $V(r)=-V_0 r^{2q}\exp{(-q(r^2-1))}$,  $\nu_s=\textrm{sgn} (c_0)=+1$, and  $\nu_a=c_2/c_0=-4.66\cdot 10^{-3}$
for a $^{87}$Rb condensate. The total number of atoms in the experiment of Ref.~\cite{Beattie13} is $N=8\cdot 10^4$, the parameters of the trap are  $\omega_z=350$Hz and $R=12\mu$m.
The dimensionless parameters in our model are chosen to correspond to the conditions of the experiment: depth of the potential $ V_0/(\hbar\omega_r)=1574.3 $, quadratic Zeeman shift $\epsilon= \delta E/(\hbar\omega_r)=9070.4.$

Non-equilibrium dissipative effects are of crucial importance since they provide the mechanism for the damping of the vortices.
We describe such effects by the phenomenological parameter $\gamma>0$ on the left hand side of GPEs (\ref{main1}) and (\ref{main2}).
Dissipation means that energy and particles are exchanged between the condensate and a thermal reservoir. Here, we use the phenomenological approach introduced by Choi et al. \cite{Choi98} for atomic BEC in a manner similar to that originally proposed by Pitaevskii \cite{Pitaevskii59}. Dissipative GPEs are used extensively in studies of vortex dynamics \cite{Tsubota03,Tsubota13,PhysRevA.77.023605}.

The parameter $\gamma$  may be calculated using quantum kinetic theory \cite{PRA62.033606} for homogeneous BEC. In the following, we neglect a possible position and temperature dependence of $\gamma$, and set  $\gamma=0.08$.
With this value for the parameter $\gamma$, we found the vortex lifetime  obtained from numerical simulations in a simply connected sheet trap to be in agreement with results of experimental studies of the dissipative vortex dynamics \cite{CavendishSingleComponent}. Furthermore, we verified that our main results do not depend qualitatively on the chosen value of $\gamma\ll 1$.

The GPEs (\ref{main1}) and (\ref{main2}) with $\gamma>0$ conserve neither the energy nor the number of particles. In a study of  hydrodynamic quantum turbulence~\cite{Tsubota13} a time-dependent chemical potential $\mu(t)$  was introduced, so that the number of particles was conserved.
For a relatively rapid process, such as vortex nucleation or decay, the specific time-dependence of the chemical potential is not very important as was pointed out in Ref.~\cite{Tsubota03}. Here, we are interested in the dynamics of a stable super-flow, the lifetime of which is restricted only by the decay of the condensate itself. Therefore, in our simulations, we adjust the chemical potential $\mu(t)$ at each time step so that the  number of condensed particles slowly decays with time, $N(t)=N(0)e^{- t/\tau_0}$, with $\tau_0=180$ in order to match measurements for lifetime of atomic BECs reported in Ref.~\cite{Beattie13}.

\begin{figure}
  \includegraphics[width=3.4in]{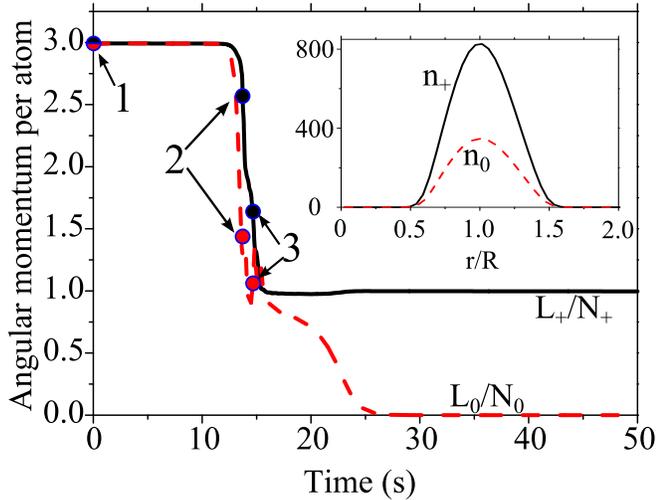}
  \caption{(Color online) Decay of the persistent current in a two-component condensate with spin-population imbalance $P_z=(N_+-N_0)/(N_++N_0)=0.4$. Shown are the angular momentum per atom for spin component $m_F=+1$ (solid black curve) and $m_F=0$ (dashed red curve).  The inset represents the initial radial distributions of the 2D densities $|\psi_+|^2$ and $|\psi_0|^2$ for the triply charged persistent current in the toroidal trap.}
  \label{angularMomentum}
\end{figure}

\begin{figure}
  \includegraphics[width=3.4in]{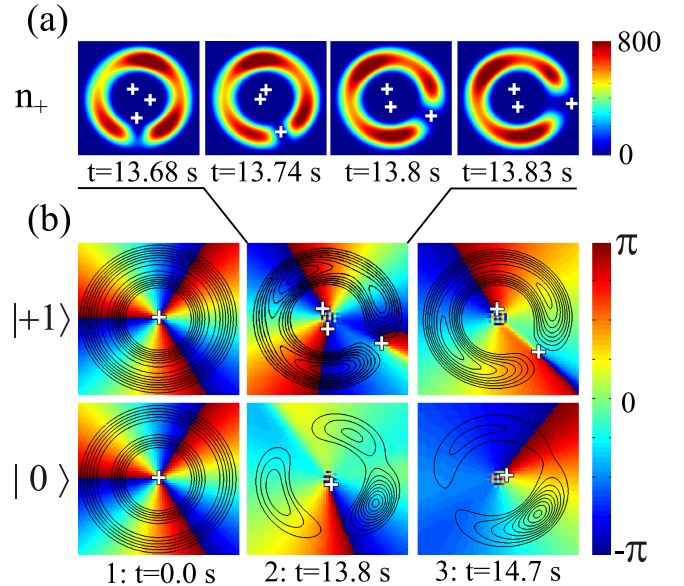}
  \caption{(Color online) (a) The detailed dynamics of a phase slip from charge-$3$ vortex to a charge-2 wortex. The colour code shows the density of the $m_F=+1$ spin component. The positions of the vortex cores are indicated by crosses. (b) Color-coded phase of two spin components combined with density isolines during phase-slips at the times indicated in Fig. \ref{angularMomentum} by the integers.}
  \label{phase}
\end{figure}

\textit{Dynamics of phase-slips} --
We numerically simulate the dynamics of the persistent flow in a toroidal trap using the two-dimensional dissipative mean-field model given by Eqs.~(\ref{main1}) and (\ref{main2}). As initial condition $\vec\psi|_{t=0}=(\psi_+,\psi_0,0)$ we employ a stationary vortex state obtained from Eqs. (\ref{main1}) and (\ref{main2}) by imaginary time propagation (see, for example, the inset in Fig. \ref{angularMomentum}). Although in the initial state the $\psi_-$ component is set to zero in accordance with the experimental conditions~\cite{Beattie13}, we allow population of this component during evolution. It turns out that due to the weakness of the spin-dependent part of the interaction and a rather significant dissipation, the population $N_-$ remains very low and can be neglected.

The angular momentum per particle is quantized for stable persistent currents in accordance with the quantization condition (\ref{quantization}). During a phase slip the angular momentum is not integer, as is clearly seen from Fig. \ref{angularMomentum} which presents a typical example of the temporal dynamics of $m_F=+1$ and $m_F=0$ spin components of the order parameter.

We observe that during a phase slip the density distribution of the two spin components is highly anisotropic
(see Fig.~\ref{phase}(a)). It turns out that the smaller the admixture of the minor component the more time is needed for the development of this symmetry-breaking azimuthal instability, and thus the longer is the lifetime of the persistent current. This is no surprise, since the phase separation appears as a result of a repulsive nonlinear inter-component interaction. The phase separation grows out of small azimuthal perturbations which are energetically favored in a condensate with comparable number of atoms in both components.

It is obvious from Fig. \ref{phase}(b) that as the result of the phase separation, the regions with reduced density in one component are filled by atoms of the other component (so that the total density distribution $n=n_++n_0$ remains axially-symmetric). These dynamical weak links play a key role for the decay of the persistent current: The vortex line, which is trapped inside the internal toroidal hole, can easily drift through these `gates', which consequently change the topological charge by one unit.

This hypothesis is supported by investigations of the dynamics of the phase of the order parameter and by analysis of the vortex core position during the phase slip. Typical examples of phase-slips are shown in Fig. \ref{phase}, where the white crosses indicate the positions of the vortex cores.
For the detection of the vortex core position we used a simple phase unwrapping technique similar to the one described in Ref. \cite{Caradoc-DaviesPhD}.
In low-density regions near the center of the toroidal trap one finds a sea of ``ghost" vortices, which correspond to strong phase fluctuations well outside of the Thomas-Fermi surface.

The physical nature of the phase slips, which are observed in our simulations, is similar to the mechanics of the decay of superflow in toroidal traps with weak links.  They have been investigated both experimentally~\cite{Ramanathan11,Wright13} and theoretically~\cite{Brand01,Piazza09, Piazza2013}. Indeed, the density variations in each component, which arise dynamically in toroidal spinor BECs as the result of an instability, play the same role in the decay of the persistent currents as the localized regions of reduced superfluid density  produced by the external barrier placed across the annulus. It was found  \cite{Piazza09} that in the region where the condensate density is  reduced by the barrier, a vortex line moves from the inner region and either circulates around the annulus or collides and annihilates with an anti-vortex, which enters the toroid from outside the annulus. As a result, the condensate suffers a global $2\pi$ phase slip and the total angular momentum decreases by one unit. It turns out that dissipative effects dramatically change the dynamics of the vortex lines during the phase slip: the circulation around the annulus of the vortex line becomes unstable in the condensate with dissipation. The phase slip appears as a result of the fast drift of the vortex line through the dynamical weak links to the edge of the condensate, where the vortex decays into elementary excitations. This mechanism  of the decay of the persistent flow appears to be significantly more effective, than the vortex-antivortex collision and annihilation inside the weak link, which plays a crucial role in dissipationless condensates~\cite{Piazza09,Piazza2013}.

\begin{figure}
  \includegraphics[width=90mm]{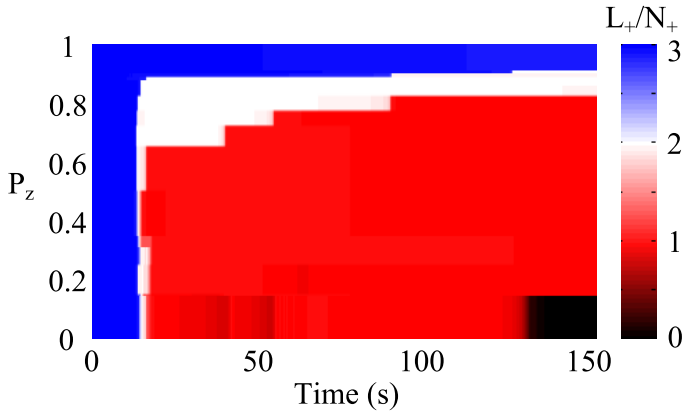}
  \caption{(Color online) Angular momentum per atom $L_+/N_+$ as a function of time and $P_z$. Angular momentum is represented by the color code and characterizes vortex charge if $L_+/N_+$ is integer. It can be seen that there is a very sharp edge between stable and unstable currents at about $P_z\approx0.89$.}
  \label{Pz}
\end{figure}

To compare our theoretical findings with the experiment \cite{Beattie13} we present the angular momentum per particle in the major component as the function of the spin-population imbalance $P_z=(N_+-N_0)/(N_++N_0)$ and time. In agreement with the experiments it is seen from Fig. \ref{Pz} that above a well-defined (`critical') value for $P_z$ the persistent current is stable. Below this critical value the supercurrent decays rapidly. However, the
experimentally observed `critical' value ($P_z\approx 0.64$) is well below our predictions ($P_z\approx 0.89$). The difference between our theoretical estimates and experimental results could be minimized, if one properly accounts for possible deviation in the experimental conditions \cite{CavendishSingleComponent} of 3D distribution of the trapping potential from the pure Laguerre-Gauss shape assumed in our 2D model.   A more accurate description of the dissipative processes could also improve the quantitative agreement with the experiment. Indeed, the phenomenological parameter $\gamma$, which describes the dissipation in our simulations, is in general temperature- and position-dependent, which is not included in our simple model.

\textit{Conclusions} --
We studied the stability of persistent flow in spinor Bose-Einstein condensates within a dissipative mean field theory. We find that in a BEC with dominant population in one spin component, persistent currents are stable for over two minutes. Increasing the population of atoms in a second spin component, we observe that at a rather well-defined `critical' population imbalance the persistent flow gets unstable and quickly looses vorticity in well defined steps.

We numerically simulate the dynamics of this quantized loss of vorticity. Using phase unwrapping of the condensate wave function we detect the position of the vortex cores. Thus, we prove that the single-charge vortex line enters into the torus from the inner region through a `gate' - region in the condensate annulus with reduced density of one component filled by atoms of the other component. This local phase separation of the spin components occurs dynamically because of the nonlinear inter-component interactions. Due to dissipation, which is phenomenologically implemented into our model, the vortex line then quickly traverses these self-induced weak links from the dense condensate region to the edge of the toroidal cloud, where it immediately decays into elementary excitations. During this process the vorticity slips by one unit. Such quantized phase slips occur until the persistent flow stops.

Our observations qualitatively agree with the experimental findings reported in Ref.~\cite{Beattie13}.
We chose parameters for our calculations such that they closely resemble the conditions
of this experiment. However, the  mechanism of the persistent current decay suggested in the present work
most probably can be similarly observed in any multi-component condensate with repulsive interatomic interactions.

\textit{Acknowledgment} --
This work was supported by the Deutsche Akade\-mische Austauschdienst (DAAD),
Germany, and the State Fund for Fundamental Researches, Ukraine.

\bibliography{toroidal2}

\begin{thebibliography}{29}
\expandafter\ifx\csname natexlab\endcsname\relax\def\natexlab#1{#1}\fi
\expandafter\ifx\csname bibnamefont\endcsname\relax
  \def\bibnamefont#1{#1}\fi
\expandafter\ifx\csname bibfnamefont\endcsname\relax
  \def\bibfnamefont#1{#1}\fi
\expandafter\ifx\csname citenamefont\endcsname\relax
  \def\citenamefont#1{#1}\fi
\expandafter\ifx\csname url\endcsname\relax
  \def\url#1{\texttt{#1}}\fi
\expandafter\ifx\csname urlprefix\endcsname\relax\def\urlprefix{URL }\fi
\providecommand{\bibinfo}[2]{#2}
\providecommand{\eprint}[2][]{\url{#2}}

\bibitem[{\citenamefont{{Ryu} et~al.}(2007)\citenamefont{{Ryu}, {Andersen},
  {Clad{\'e}}, {Natarajan}, {Helmerson}, and {Phillips}}}]{Ryu07}
\bibinfo{author}{\bibfnamefont{C.}~\bibnamefont{{Ryu}}},
  \bibinfo{author}{\bibfnamefont{M.~F.} \bibnamefont{{Andersen}}},
  \bibinfo{author}{\bibfnamefont{P.}~\bibnamefont{{Clad{\'e}}}},
  \bibinfo{author}{\bibfnamefont{V.}~\bibnamefont{{Natarajan}}},
  \bibinfo{author}{\bibfnamefont{K.}~\bibnamefont{{Helmerson}}},
  \bibnamefont{and} \bibinfo{author}{\bibfnamefont{W.~D.}
  \bibnamefont{{Phillips}}}, \bibinfo{journal}{\prl}
  \textbf{\bibinfo{volume}{99}}, \bibinfo{eid}{260401} (\bibinfo{year}{2007}).

\bibitem[{\citenamefont{{Ramanathan} et~al.}(2011)\citenamefont{{Ramanathan},
  {Wright}, {Muniz}, {Zelan}, {Hill}, {Lobb}, {Helmerson}, {Phillips}, and
  {Campbell}}}]{Ramanathan11}
\bibinfo{author}{\bibfnamefont{A.}~\bibnamefont{{Ramanathan}}},
  \bibinfo{author}{\bibfnamefont{K.~C.} \bibnamefont{{Wright}}},
  \bibinfo{author}{\bibfnamefont{S.~R.} \bibnamefont{{Muniz}}},
  \bibinfo{author}{\bibfnamefont{M.}~\bibnamefont{{Zelan}}},
  \bibinfo{author}{\bibfnamefont{W.~T.} \bibnamefont{{Hill}},
  \bibfnamefont{III}}, \bibinfo{author}{\bibfnamefont{C.~J.}
  \bibnamefont{{Lobb}}},
  \bibinfo{author}{\bibfnamefont{K.}~\bibnamefont{{Helmerson}}},
  \bibinfo{author}{\bibfnamefont{W.~D.} \bibnamefont{{Phillips}}},
  \bibnamefont{and} \bibinfo{author}{\bibfnamefont{G.~K.}
  \bibnamefont{{Campbell}}}, \bibinfo{journal}{\prl}
  \textbf{\bibinfo{volume}{106}}, \bibinfo{eid}{130401} (\bibinfo{year}{2011}).

\bibitem[{\citenamefont{Moulder et~al.}(2012)\citenamefont{Moulder, Beattie,
  Smith, Tammuz, and Hadzibabic}}]{CavendishSingleComponent}
\bibinfo{author}{\bibfnamefont{S.}~\bibnamefont{Moulder}},
  \bibinfo{author}{\bibfnamefont{S.}~\bibnamefont{Beattie}},
  \bibinfo{author}{\bibfnamefont{R.~P.} \bibnamefont{Smith}},
  \bibinfo{author}{\bibfnamefont{N.}~\bibnamefont{Tammuz}}, \bibnamefont{and}
  \bibinfo{author}{\bibfnamefont{Z.}~\bibnamefont{Hadzibabic}},
  \bibinfo{journal}{Phys. Rev. A} \textbf{\bibinfo{volume}{86}},
  \bibinfo{pages}{013629} (\bibinfo{year}{2012}).

\bibitem[{\citenamefont{Wright et~al.}(2013)\citenamefont{Wright, Blakestad,
  Lobb, Phillips, and Campbell}}]{Wright13}
\bibinfo{author}{\bibfnamefont{K.~C.} \bibnamefont{Wright}},
  \bibinfo{author}{\bibfnamefont{R.~B.} \bibnamefont{Blakestad}},
  \bibinfo{author}{\bibfnamefont{C.~J.} \bibnamefont{Lobb}},
  \bibinfo{author}{\bibfnamefont{W.~D.} \bibnamefont{Phillips}},
  \bibnamefont{and} \bibinfo{author}{\bibfnamefont{G.~K.}
  \bibnamefont{Campbell}}, \bibinfo{journal}{Phys. Rev. Lett.}
  \textbf{\bibinfo{volume}{110}}, \bibinfo{pages}{025302}
  (\bibinfo{year}{2013}).

\bibitem[{\citenamefont{Beattie et~al.}(2013)\citenamefont{Beattie, Moulder,
  Fletcher, and Hadzibabic}}]{Beattie13}
\bibinfo{author}{\bibfnamefont{S.}~\bibnamefont{Beattie}},
  \bibinfo{author}{\bibfnamefont{S.}~\bibnamefont{Moulder}},
  \bibinfo{author}{\bibfnamefont{R.~J.} \bibnamefont{Fletcher}},
  \bibnamefont{and}
  \bibinfo{author}{\bibfnamefont{Z.}~\bibnamefont{Hadzibabic}},
  \bibinfo{journal}{Phys. Rev. Lett.} \textbf{\bibinfo{volume}{110}},
  \bibinfo{pages}{025301} (\bibinfo{year}{2013}).

\bibitem[{\citenamefont{{Kanamoto} et~al.}(2009)\citenamefont{{Kanamoto},
  {Carr}, and {Ueda}}}]{KanamotoUedaPRA09}
\bibinfo{author}{\bibfnamefont{R.}~\bibnamefont{{Kanamoto}}},
  \bibinfo{author}{\bibfnamefont{L.~D.} \bibnamefont{{Carr}}},
  \bibnamefont{and} \bibinfo{author}{\bibfnamefont{M.}~\bibnamefont{{Ueda}}},
  \bibinfo{journal}{\pra} \textbf{\bibinfo{volume}{79}}, \bibinfo{eid}{063616}
  (\bibinfo{year}{2009}).

\bibitem[{\citenamefont{{Kanamoto} et~al.}(2010)\citenamefont{{Kanamoto},
  {Carr}, and {Ueda}}}]{KanamotoUedaPRA10}
\bibinfo{author}{\bibfnamefont{R.}~\bibnamefont{{Kanamoto}}},
  \bibinfo{author}{\bibfnamefont{L.~D.} \bibnamefont{{Carr}}},
  \bibnamefont{and} \bibinfo{author}{\bibfnamefont{M.}~\bibnamefont{{Ueda}}},
  \bibinfo{journal}{\pra} \textbf{\bibinfo{volume}{81}}, \bibinfo{eid}{023625}
  (\bibinfo{year}{2010}).

\bibitem[{\citenamefont{{Yulin} et~al.}(2011)\citenamefont{{Yulin}, {Bludov},
  {Konotop}, {Kuzmiak}, and {Salerno}}}]{YulinPRA11}
\bibinfo{author}{\bibfnamefont{A.~V.} \bibnamefont{{Yulin}}},
  \bibinfo{author}{\bibfnamefont{Y.~V.} \bibnamefont{{Bludov}}},
  \bibinfo{author}{\bibfnamefont{V.~V.} \bibnamefont{{Konotop}}},
  \bibinfo{author}{\bibfnamefont{V.}~\bibnamefont{{Kuzmiak}}},
  \bibnamefont{and}
  \bibinfo{author}{\bibfnamefont{M.}~\bibnamefont{{Salerno}}},
  \bibinfo{journal}{\pra} \textbf{\bibinfo{volume}{84}}, \bibinfo{eid}{063638}
  (\bibinfo{year}{2011}).

\bibitem[{\citenamefont{Mason and Berloff}(2009)}]{Berloff09}
\bibinfo{author}{\bibfnamefont{P.}~\bibnamefont{Mason}} \bibnamefont{and}
  \bibinfo{author}{\bibfnamefont{N.~G.} \bibnamefont{Berloff}},
  \bibinfo{journal}{Phys. Rev. A} \textbf{\bibinfo{volume}{79}},
  \bibinfo{pages}{043620} (\bibinfo{year}{2009}).

\bibitem[{\citenamefont{{Smyrnakis} et~al.}(2009)\citenamefont{{Smyrnakis},
  {Bargi}, {Kavoulakis}, {Magiropoulos}, {K{\"a}rkk{\"a}inen}, and
  {Reimann}}}]{SmyrnakisPRL09}
\bibinfo{author}{\bibfnamefont{J.}~\bibnamefont{{Smyrnakis}}},
  \bibinfo{author}{\bibfnamefont{S.}~\bibnamefont{{Bargi}}},
  \bibinfo{author}{\bibfnamefont{G.~M.} \bibnamefont{{Kavoulakis}}},
  \bibinfo{author}{\bibfnamefont{M.}~\bibnamefont{{Magiropoulos}}},
  \bibinfo{author}{\bibfnamefont{K.}~\bibnamefont{{K{\"a}rkk{\"a}inen}}},
  \bibnamefont{and} \bibinfo{author}{\bibfnamefont{S.~M.}
  \bibnamefont{{Reimann}}}, \bibinfo{journal}{\prl}
  \textbf{\bibinfo{volume}{103}}, \bibinfo{eid}{100404} (\bibinfo{year}{2009}).

\bibitem[{\citenamefont{{Smyrnakis} et~al.}(2010)\citenamefont{{Smyrnakis},
  {Magiropoulos}, {Kavoulakis}, and {Jackson}}}]{SmyrnakisPRA10}
\bibinfo{author}{\bibfnamefont{J.}~\bibnamefont{{Smyrnakis}}},
  \bibinfo{author}{\bibfnamefont{M.}~\bibnamefont{{Magiropoulos}}},
  \bibinfo{author}{\bibfnamefont{G.~M.} \bibnamefont{{Kavoulakis}}},
  \bibnamefont{and} \bibinfo{author}{\bibfnamefont{A.~D.}
  \bibnamefont{{Jackson}}}, \bibinfo{journal}{\pra}
  \textbf{\bibinfo{volume}{81}}, \bibinfo{eid}{063601} (\bibinfo{year}{2010}).

\bibitem[{\citenamefont{{Malet} et~al.}(2010)\citenamefont{{Malet},
  {Kavoulakis}, and {Reimann}}}]{SmyrnakisPRA10_annular}
\bibinfo{author}{\bibfnamefont{F.}~\bibnamefont{{Malet}}},
  \bibinfo{author}{\bibfnamefont{G.~M.} \bibnamefont{{Kavoulakis}}},
  \bibnamefont{and} \bibinfo{author}{\bibfnamefont{S.~M.}
  \bibnamefont{{Reimann}}}, \bibinfo{journal}{\pra}
  \textbf{\bibinfo{volume}{81}}, \bibinfo{eid}{013630} (\bibinfo{year}{2010}).

\bibitem[{\citenamefont{{K{\"a}rkk{\"a}inen}
  et~al.}(2007)\citenamefont{{K{\"a}rkk{\"a}inen}, {Christensson}, {Reinisch},
  {Kavoulakis}, and {Reimann}}}]{SmyrnakisPRA07}
\bibinfo{author}{\bibfnamefont{K.}~\bibnamefont{{K{\"a}rkk{\"a}inen}}},
  \bibinfo{author}{\bibfnamefont{J.}~\bibnamefont{{Christensson}}},
  \bibinfo{author}{\bibfnamefont{G.}~\bibnamefont{{Reinisch}}},
  \bibinfo{author}{\bibfnamefont{G.~M.} \bibnamefont{{Kavoulakis}}},
  \bibnamefont{and} \bibinfo{author}{\bibfnamefont{S.~M.}
  \bibnamefont{{Reimann}}}, \bibinfo{journal}{\pra}
  \textbf{\bibinfo{volume}{76}}, \bibinfo{eid}{043627} (\bibinfo{year}{2007}).

\bibitem[{\citenamefont{{Jackson} and
  {Kavoulakis}}(2006)}]{JacksonKavoulakisPRA06}
\bibinfo{author}{\bibfnamefont{A.~D.} \bibnamefont{{Jackson}}}
  \bibnamefont{and} \bibinfo{author}{\bibfnamefont{G.~M.}
  \bibnamefont{{Kavoulakis}}}, \bibinfo{journal}{\pra}
  \textbf{\bibinfo{volume}{74}}, \bibinfo{eid}{065601} (\bibinfo{year}{2006}).

\bibitem[{\citenamefont{{{\"O}gren} and {Kavoulakis}}(2007)}]{KavoulakisJLTP07}
\bibinfo{author}{\bibfnamefont{M.}~\bibnamefont{{{\"O}gren}}} \bibnamefont{and}
  \bibinfo{author}{\bibfnamefont{G.~M.} \bibnamefont{{Kavoulakis}}},
  \bibinfo{journal}{Journal of Low Temperature Physics}
  \textbf{\bibinfo{volume}{149}}, \bibinfo{pages}{176} (\bibinfo{year}{2007}).

\bibitem[{\citenamefont{{Brtka} et~al.}(2010)\citenamefont{{Brtka}, {Gammal},
  and {Malomed}}}]{BrtkaMalomedPRA10}
\bibinfo{author}{\bibfnamefont{M.}~\bibnamefont{{Brtka}}},
  \bibinfo{author}{\bibfnamefont{A.}~\bibnamefont{{Gammal}}}, \bibnamefont{and}
  \bibinfo{author}{\bibfnamefont{B.~A.} \bibnamefont{{Malomed}}},
  \bibinfo{journal}{\pra} \textbf{\bibinfo{volume}{82}}, \bibinfo{eid}{053610}
  (\bibinfo{year}{2010}).

\bibitem[{\citenamefont{Anoshkin et~al.}(2013)\citenamefont{Anoshkin, Wu, and
  Zaremba}}]{ANO13}
\bibinfo{author}{\bibfnamefont{K.}~\bibnamefont{Anoshkin}},
  \bibinfo{author}{\bibfnamefont{Z.}~\bibnamefont{Wu}}, \bibnamefont{and}
  \bibinfo{author}{\bibfnamefont{E.}~\bibnamefont{Zaremba}},
  \bibinfo{journal}{Phys. Rev. A} \textbf{\bibinfo{volume}{88}},
  \bibinfo{pages}{013609} (\bibinfo{year}{2013}).

\bibitem[{\citenamefont{Bargi et~al.}(2010)\citenamefont{Bargi, Malet,
  Kavoulakis, and Reimann}}]{BargiKavoulakis10}
\bibinfo{author}{\bibfnamefont{S.}~\bibnamefont{Bargi}},
  \bibinfo{author}{\bibfnamefont{F.}~\bibnamefont{Malet}},
  \bibinfo{author}{\bibfnamefont{G.~M.} \bibnamefont{Kavoulakis}},
  \bibnamefont{and} \bibinfo{author}{\bibfnamefont{S.~M.}
  \bibnamefont{Reimann}}, \bibinfo{journal}{\pra}
  \textbf{\bibinfo{volume}{82}}, \bibinfo{pages}{043631}
  (\bibinfo{year}{2010}).

\bibitem[{\citenamefont{\ifmmode~\acute{S}\else \'{S}\fi{}wis\l{}ocki and
  Matuszewski}(2012)}]{Matuszewski12}
\bibinfo{author}{\bibfnamefont{T.}~\bibnamefont{\ifmmode~\acute{S}\else
  \'{S}\fi{}wis\l{}ocki}} \bibnamefont{and}
  \bibinfo{author}{\bibfnamefont{M.}~\bibnamefont{Matuszewski}},
  \bibinfo{journal}{Phys. Rev. A} \textbf{\bibinfo{volume}{85}},
  \bibinfo{pages}{023601} (\bibinfo{year}{2012}).

\bibitem[{\citenamefont{Choi et~al.}(1998)\citenamefont{Choi, Morgan, and
  Burnett}}]{Choi98}
\bibinfo{author}{\bibfnamefont{S.}~\bibnamefont{Choi}},
  \bibinfo{author}{\bibfnamefont{S.~A.} \bibnamefont{Morgan}},
  \bibnamefont{and} \bibinfo{author}{\bibfnamefont{K.}~\bibnamefont{Burnett}},
  \bibinfo{journal}{Phys. Rev. A} \textbf{\bibinfo{volume}{57}},
  \bibinfo{pages}{4057} (\bibinfo{year}{1998}).

\bibitem[{\citenamefont{Pitaevskii}(1959)}]{Pitaevskii59}
\bibinfo{author}{\bibfnamefont{L.}~\bibnamefont{Pitaevskii}},
  \bibinfo{journal}{Sov. Phys. JETP} \textbf{\bibinfo{volume}{8}},
  \bibinfo{pages}{88} (\bibinfo{year}{1959}).

\bibitem[{\citenamefont{Kasamatsu et~al.}(2003)\citenamefont{Kasamatsu,
  Tsubota, and Ueda}}]{Tsubota03}
\bibinfo{author}{\bibfnamefont{K.}~\bibnamefont{Kasamatsu}},
  \bibinfo{author}{\bibfnamefont{M.}~\bibnamefont{Tsubota}}, \bibnamefont{and}
  \bibinfo{author}{\bibfnamefont{M.}~\bibnamefont{Ueda}},
  \bibinfo{journal}{Phys. Rev. A} \textbf{\bibinfo{volume}{67}},
  \bibinfo{pages}{033610} (\bibinfo{year}{2003}).

\bibitem[{\citenamefont{{Tsubota} et~al.}(2013)\citenamefont{{Tsubota},
  {Kobayashi}, and {Takeuchi}}}]{Tsubota13}
\bibinfo{author}{\bibfnamefont{M.}~\bibnamefont{{Tsubota}}},
  \bibinfo{author}{\bibfnamefont{M.}~\bibnamefont{{Kobayashi}}},
  \bibnamefont{and}
  \bibinfo{author}{\bibfnamefont{H.}~\bibnamefont{{Takeuchi}}},
  \bibinfo{journal}{physrep} \textbf{\bibinfo{volume}{522}},
  \bibinfo{pages}{191} (\bibinfo{year}{2013}).

\bibitem[{\citenamefont{Carretero-Gonz\'alez
  et~al.}(2008)\citenamefont{Carretero-Gonz\'alez, Whitaker, Kevrekidis, and
  Frantzeskakis}}]{PhysRevA.77.023605}
\bibinfo{author}{\bibfnamefont{R.}~\bibnamefont{Carretero-Gonz\'alez}},
  \bibinfo{author}{\bibfnamefont{N.}~\bibnamefont{Whitaker}},
  \bibinfo{author}{\bibfnamefont{P.~G.} \bibnamefont{Kevrekidis}},
  \bibnamefont{and} \bibinfo{author}{\bibfnamefont{D.~J.}
  \bibnamefont{Frantzeskakis}}, \bibinfo{journal}{Phys. Rev. A}
  \textbf{\bibinfo{volume}{77}}, \bibinfo{pages}{023605}
  (\bibinfo{year}{2008}).

\bibitem[{\citenamefont{{Lee} and {Gardiner}}(2000)}]{PRA62.033606}
\bibinfo{author}{\bibfnamefont{M.~D.} \bibnamefont{{Lee}}} \bibnamefont{and}
  \bibinfo{author}{\bibfnamefont{C.~W.} \bibnamefont{{Gardiner}}},
  \bibinfo{journal}{\pra} \textbf{\bibinfo{volume}{62}}, \bibinfo{eid}{033606}
  (\bibinfo{year}{2000}).

\bibitem[{\citenamefont{{Caradoc-Davies}}(2000)}]{Caradoc-DaviesPhD}
\bibinfo{author}{\bibfnamefont{B.}~\bibnamefont{{Caradoc-Davies}}},
  \emph{\bibinfo{title}{{Vortex dynamics in Bose-Einstein condensate, Ph.D.
  thesis}}} (\bibinfo{publisher}{University of Otago}, \bibinfo{address}{(NZ)},
  \bibinfo{year}{2000}).

\bibitem[{\citenamefont{{Brand} and {Reinhardt}}(2001)}]{Brand01}
\bibinfo{author}{\bibfnamefont{J.}~\bibnamefont{{Brand}}} \bibnamefont{and}
  \bibinfo{author}{\bibfnamefont{W.~P.} \bibnamefont{{Reinhardt}}},
  \bibinfo{journal}{Journal of Physics B Atomic Molecular Physics}
  \textbf{\bibinfo{volume}{34}}, \bibinfo{pages}{L113} (\bibinfo{year}{2001}).

\bibitem[{\citenamefont{{Piazza} et~al.}(2009)\citenamefont{{Piazza},
  {Collins}, and {Smerzi}}}]{Piazza09}
\bibinfo{author}{\bibfnamefont{F.}~\bibnamefont{{Piazza}}},
  \bibinfo{author}{\bibfnamefont{L.~A.} \bibnamefont{{Collins}}},
  \bibnamefont{and} \bibinfo{author}{\bibfnamefont{A.}~\bibnamefont{{Smerzi}}},
  \bibinfo{journal}{\pra} \textbf{\bibinfo{volume}{80}}, \bibinfo{eid}{021601}
  (\bibinfo{year}{2009}).

\bibitem[{\citenamefont{{Piazza} et~al.}(2013)\citenamefont{{Piazza},
  {Collins}, and {Smerzi}}}]{Piazza2013}
\bibinfo{author}{\bibfnamefont{F.}~\bibnamefont{{Piazza}}},
  \bibinfo{author}{\bibfnamefont{L.~A.} \bibnamefont{{Collins}}},
  \bibnamefont{and} \bibinfo{author}{\bibfnamefont{A.}~\bibnamefont{{Smerzi}}},
  \bibinfo{journal}{Journal of Physics B Atomic Molecular Physics}
  \textbf{\bibinfo{volume}{46}}, \bibinfo{eid}{095302} (\bibinfo{year}{2013}).

\end{thebibliography}
\end{document}